\documentclass[12pt]{article}
\usepackage{amsmath}
\usepackage{graphics,graphicx,epsfig}
\usepackage{amssymb}
\usepackage{epstopdf}
\usepackage{sidecap}
\usepackage{appendix}

\def\tildet{\tilde T}

\begin{document}

\begin{titlepage}
\vfill
\begin{flushright}
ACFI-T15-06
\end{flushright}

\vfill
\begin{center}
\baselineskip=16pt
{\Large\bf Melvin Magnetic Fluxtube/Cosmology Correspondence}
%{\Large\bf Anisotropic Melvin Cosmologies}
%{\Large\bf Anisotropic Magnetic Cosmologies from}
%\vskip 0.1in
%{\Large\bf Melvin Fluxtubes}
\vskip 0.15in
\vskip 10.mm
{\large\bf 
David Kastor\footnote{kastor@physics.umass.edu} and Jennie Traschen\footnote{traschen@physics.umass.edu}} 

\vskip 0.5cm
{{Amherst Center for Fundamental Interactions\\
Department of Physics, University of Massachusetts, Amherst, MA 01003
     }}
\vspace{6pt}
\end{center}
\vskip 0.2in
\par
\begin{center}
{\bf Abstract}
 \end{center}
\begin{quote}
We explore a correspondence  between Melvin magnetic fluxtubes and anisotropic cosmological solutions, which we call `Melvin cosmologies'.  The correspondence via analytic continuation provides useful information in both directions.  Solution generating techniques known on the fluxtube side can also be used for generating cosmological backgrounds.  Melvin cosmologies interpolate between different limiting Kasner behaviors at early and late times.  This has an analogue on the fluxtube side between limiting Levi-Civita behavior at small and large radii.    We construct generalized Melvin fluxtubes and cosmologies in both Einstein-Maxwell theory and dilaton gravity and show that similar properties hold.

\vfill
\vskip 2.mm
\end{quote}
\hfill
\end{titlepage}

%\tableofcontents

\section{Introduction}

It is often fruitful to explore underlying connections that exist between different types of physical phenomena, the AdS/CFT correspondence being a prominent example.  A related example, known as the 
the domain wall/cosmology correspondence was introduced in  \cite{Skenderis:2006jq,Skenderis:2006rr,Skenderis:2006fb}.  The simplest instance of this correspondence is the map from AdS, regarded as a domain wall, to dS, regarded as a cosmology, via analytic continuation. 
Starting with AdS written in Poincare coordinates with cosmological constant $\Lambda=-3/l^2$
\begin{equation}\label{poincare}
ds^2={l^2\over r^2}(dr^2-dt^2 +dy^2+dz^2),
\end{equation}
one views the $t$, $y$ and $z$ coordinates as tangent to a static, domain wall with evolution in the radial direction $r$.
Slices tangent to the wall, {\it i.e.} at constant $r$, are Poincare invariant.
Under the analytic continuation $r=iT$, $t=ix$, $l=iL$, these constant $r$ slices map onto constant $T$ slices of dS  in cosmological coordinates with flat spatial sections with cosmological constant 
$\Lambda = +3/L^2$
\begin{equation}\label{poindual}
ds^2 = {L^2\over T^2}(-dT^2 +dx^2+dy^2 +dz^2)
\end{equation}
The Poincare invariance of the AdS constant $r$ slices  maps to the homogeneity and isotropy of the dS constant $T$ slices, and vice-versa.

In this paper we will study a similar connection between domain wall and cosmological spacetimes, but with a few key differences.  First, rather than a cosmological constant,which provides curvature, we will consider spacetimes with electromagnetic fields.   Second, our domain walls and cosmologies will be anisotropic, with an electric or magnetic field providing a preferred spatial direction.  The domain walls  we consider are generalizations of the Melvin spacetime\footnote{As we discuss below, Melvin \cite{Melvin:1963qx} may be regarded as a magnetic fluxtube, as it most commonly is, or alternatively as an anisotropic domain walls, depending on the range of coordinates chosen.  We will move back and forth, as convenient, between these two global forms of the same local solution.} \cite{Melvin:1963qx}.  The corresponding cosmologies will be generalizations of the homogeneous, but anisotropic magnetic cosmologies found by Rosen \cite{rosen1,rosen2}. 

Both types of solutions play important roles in gravitational physics.  Melvin magnetic fields  have been widely used to study phenomena such as black holes in background magnetic fields, the quantum pair creation of oppositely charged black holes, and fluxbranes in string theory.  Anisotropic magnetic cosmologies, on the other hand, have long been of interest in the context of both  primordial magnetic fields and possible large scale cosmic anisotropy. 

The generalized Melvin and anisotropic cosmological solutions we consider include both new solutions and reconstructions of existing results.  In all cases, however, the relations we find between them expose common structures and patterns.  These relations provide a bridge between different areas, allowing results in one area to be transferred directly to the other.    For example, it is well known that  solutions containing Melvin-type fields may be constructed via solution generating techniques starting from vacuum solutions \cite{harrison,Ernst3}.    The corresponding cosmological solutions have typically been found by direct solution of the field equations.  However, we will see that the solution generating methods apply in this context as well and provides additional organizational structure.   In the other direction, it was originally observed in \cite{Belinski:1973zz} that magnetic cosmologies interpolate in their evolution from early to late times between different Kasner spacetimes, which are anisotropic cosmological solutions to the vacuum field equations.  This behavior is reminiscent of the BKL oscillation between approximate Kasner solutions in the approach to a spacelike singularity \cite{Belinsky:1970ew}.  Via the correspondence of cosmologies to fluxtubes, we see that Melvin spacetimes similarly interpolate between two distinct cylindrically symmetric vacuum, Levi-Civita solutions.

The plan of this paper is as follows.  In Section (\ref{melvinsection}) we show that the Melvin magnetic fluxtube can be analytically continued to obtain an anisotropic electric cosmology, which we call the Melvin cosmology.  An alternative, magnetic field sourcing the Melvin cosmology can be obtained through electromagnetic duality.
In Section (\ref{kasnersection}) we introduce the vacuum Levi-Civita and Kasner families of spacetimes, which also form a pair under analytic continuation, and will be important in our construction of a more general class of Melvin fluxtubes and cosmologies.  In Section (\ref{interpolating}) we show that the Melvin fluxtube and cosmology act as interpolating spacetimes between respectively different vacuum Levi-Civita and Kasner limits.  In Section (\ref{generating}) we construct generalizations of the Melvin fluxtube and cosmology by applying a solution generating technique to seed Levi-Civita and Kasner spacetimes. 
 In Section (\ref{geninterpolating})  we show that these generalized Melvin solutions act as interpolators between the seed solutions and different Levi-Civita and Kasner limits and give a simple form for the interpolating map.  In Section (\ref{scalarsection}) we add a scalar dilaton field with exponential coupling to the electromagnetic field strength and show that a similar set of results hold in this theory.  In Section (\ref{discussionsection}) we offer some concluding remarks on the significance of our results and directions for future work.    We note that analytic continuation between Melvin-type fluxbranes and cosmologies has previously appeared in \cite{Gibbons:1986wg}, while the Kasner limits of Melvin cosmologies have been studied in the context of Kaluza-Klein cosmology in \cite{Wiltshire:1987ch}.
Related work has also appeared in references \cite{Miguelote:2000qi,Baykal:2005pv,Kirezli:2012vw}.

\section{Melvin fluxtube to Melvin cosmology}\label{melvinsection}

The starting point of our investigation is the well-known Melvin  solution of the Einstein-Maxwell equations \cite{Melvin:1963qx}.  The Melvin solution may alternately describe a cylindrically symmetric magnetic fluxtube or an anisotropic domain wall, depending on a choice of coordinate range.   Through analytic continuation of the domain wall, one obtains an anisotropic cosmology, which we will call the ``Melvin cosmology".
The metric and gauge potential for the original Melvin solution \cite{Melvin:1963qx} are given by
\begin{equation}\label{melvin}
ds^2 = f^2(r)(dr^2 -dt^2 +dy^2) +{r^2\over f^2(r)}d\varphi^2,\quad A={Br^2\over 2f}d\varphi, \quad  f=1+{B^2r^2\over 4}.
\end{equation}
A magnetic field points in the $y$-direction with strength proportional to the parameter $B$, which has dimensions of inverse length.  
The coordinate $\varphi$ is usually taken to be an azimuthal angle with range $0\le \varphi<2\pi$.   In this case, the total magnetic flux is  finite and given by $\Phi_B=4\pi/B$ and (\ref{melvin}) describes a static, cylindrically symmetric fluxtube, with the magnetic flux confined by its own gravitational field.
In addition to rotational symmetry in the $\varphi$-direction and  translational symmetries in the $t$ and $y$ directions, the Melvin fluxtube is also boost symmetric in the $ty$-plane.

The Melvin solution can alternatively describe an anisotropic domain wall the direction of the magnetic field giving a preferred spatial direction.
Let us rescale the dimensionless azimuthal coordinate according to $\varphi = B z$ and take the range of the dimensionful $z$ coordinate  to be 
$-\infty\le z\le \infty$.  The Melvin `domain wall' solution now has the form
\begin{equation}\label{melvin}
ds^2 = f^2(r)(dr^2 -dt^2 +dy^2) +{B^2r^2\over f^2(r)}dz^2,\quad A={B^2r^2\over 2f}dz, \quad  f=1+{B^2r^2\over 4}
\end{equation}
with the $t$, $y$ and $z$ being directions tangent to a $2+1$ dimensional wall that evolves in the radial direction $r$.
The wall is spatially homogeneous, time translation invariant, and boost invariant in the $ty$-plane.  However,  the
difference in radial dependence between the metric components $g_{yy}$ and $g_{zz}$ make it anisotropic.  The magnetic field is tangent to the wall, pointing in the $y$-direction and the total magnetic flux is infinite.

It is now straightforward to analytically continue the Melvin domain wall to obtain what we will call  the ``Melvin cosmology".  This is accomplished by setting $r=iT$, $t=ix$ and also, in order to keep the  gauge potential real, $B=-iE$, resulting in the spacetime fields
\begin{equation}\label{cosmomelvin}
ds^2 = f^2(T)(-dT^2 +dx^2+dy^2) +{E^2T^2\over f^2(T)}dz^2,\quad A={E^2 T^2\over 2f}dz,\quad 
f(T) =1+{E^2T^2\over 4},
\end{equation}
The Melvin cosmology has homogeneous, but anisotropic flat spatial slices, and
a spatially uniform electric field pointing in the $z$-direction.  The boost symmetry of the Melvin domain wall has mapped into a rotational symmetry of the Melvin cosmology  in the $xy$-plane.  
Electromagnetic duality can be used to trade the electric field for a magnetic field in the $z$-direction.  The metric is unchanged and   the resulting field strength is given by the constant 
$F_{xy}=E$.

The Melvin cosmology (\ref{cosmomelvin})  is not a new solution\footnote{Note that the Melvin cosmology is also distinct from the cosmological Melvin fluxtube solutions \cite{Kastor:2013nha}, which are both inhomogeneous and anisotropic, describing a Melvin fluxtube embedded in an FRW background.}.   As shown in Appendix (\ref{rosensolns}),
its magnetic form coincides with the plane symmetric case of a family of anisotropic magnetic cosmologies found by Rosen \cite{rosen1,rosen2}.  
The general Rosen cosmologies break all spatial rotational symmetries and this naturally leads to the question of 
whether there exist generalized Melvin domain walls, or equivalently fluxtubes,  that analytically continue into these more general anisotropic cosmologies.
Such generalized Melvin solutions can be constructed directly starting from the Rosen cosmologies via analytic continuation.  However, we will follow an alternative path  in the next sections that offers additional insight into both the domain wall/fluxtube and cosmological solutions.

\section{Kasner and Levi-Civita}\label{kasnersection}

A primary benefit from understanding underlying connections between different types of physical phenomena is the ability to translate known properties or useful techniques from one side to the other.   In the present case, it is well known that solution generating techniques can be used to generate Melvin magnetic fields starting from vacuum solutions  \cite{harrison,Ernst3}.  This not only allows one to generate new solutions, but also provides organizational structure to the space of solutions.  Cosmological solutions with magnetic fields, such as those in \cite{rosen1,rosen2}, have typically been found by direct solution of the field equations.  We will see that they can also be found using solution generating, adding to our understanding of the space of cosmological solutions.  

The starting point for these solution generating methods are vacuum solutions that share the same symmetries.  In this section we will discuss the properties of the relevant vacuum solutions.  On the fluxtube side, these are the static, cylindrically symmetric vacuum solutions found by Levi-Civita \cite{levi-civita}.  Similarly to the Melvin solution, the Levi-Civita solutions can also be taken to describe anisotropic domain  walls, by exchanging the azimuthal coordinate $\varphi$ for a Euclidean coordinate $z$.  These Levi-Civita domain walls analytically continue into Kasner, 
 anisotropic vacuum cosmologies.

\begin{figure}
\begin{center}
\includegraphics[width=0.5\textwidth]{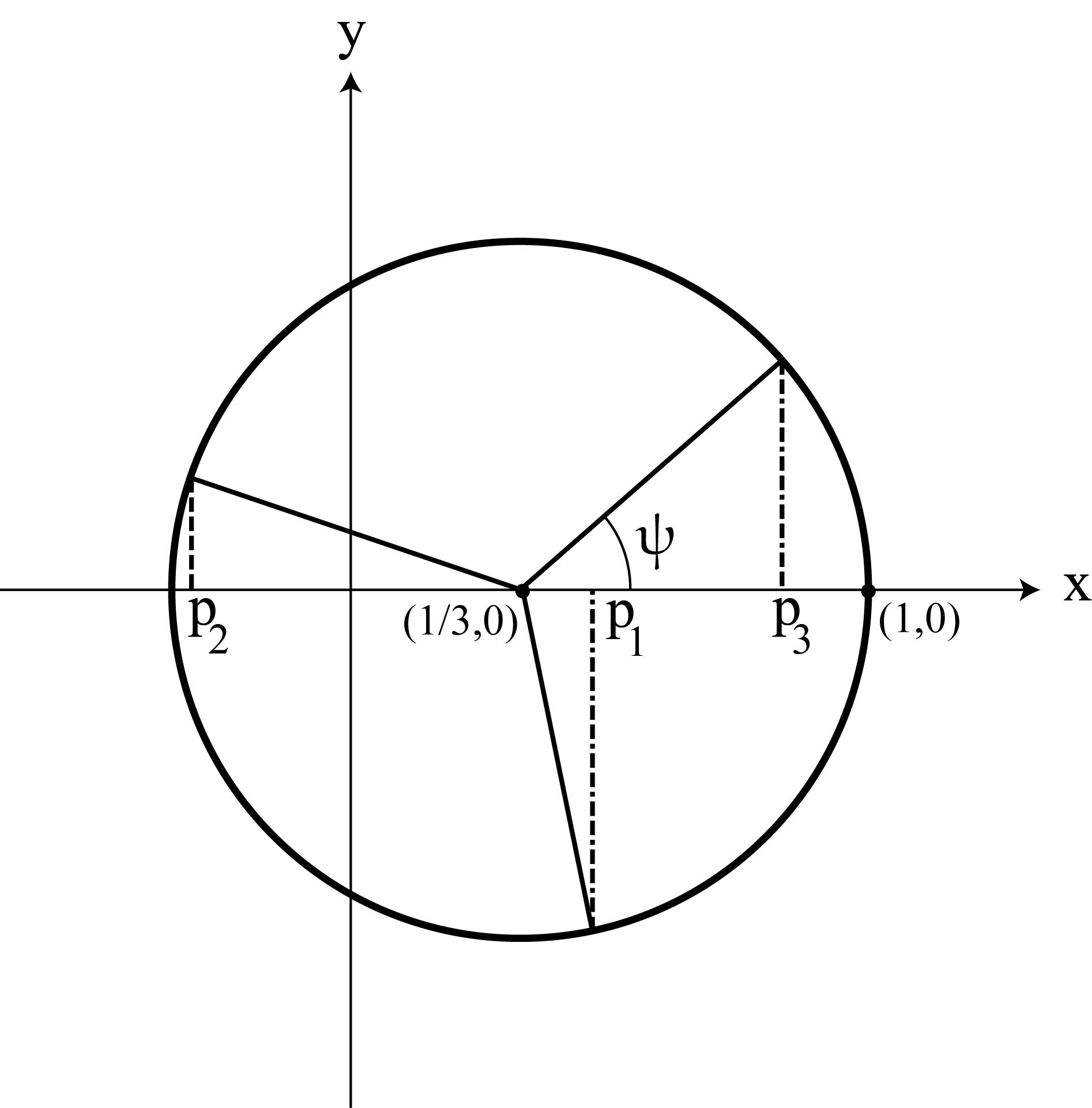}
\end{center}
\caption{{\small\sl The Kasner circle parameterizes the solutions to the Kasner constraints in terms of an angle $\psi$ and shows the relationship between the exponents $p_1$, $p_2$ and $p_3$.}} 
\label{kasner-circle}
\end{figure}

Let us  begin on the cosmology side by recalling that, while there are no isotropic, homogeneous, vacuum cosmologies with flat spatial slices, there are anisotropic solutions known as the Kasner spacetimes  \cite{Kasner:1921zz}.  These are given by
\begin{equation}\label{kasner}
%ds^2 = - dt^2 +\sum_{i=1}^3 (t/ t_0)^{2p_i}(dx^i)^2
ds^2 = - dT^2 +\sum_{i=1}^3 (T/ T_0)^{2p_i}(dx^i)^2
\end{equation}
where 
the exponents must satisfy the two constraints 
\begin{equation}\label{constraints}
\sum_{i=1}^3 p_i = \sum_{i=1}^3 (p_i)^2 =1
\end{equation}
in order to solve the vacuum Einstein equations.  Since there are $3$ Kasner exponents satisfying $2$ constraints, this leaves a $1$ parameter family of Kasner   solutions.  The parameter $T_0$ can be set to one by rescaling the spatial coordinates.   However, we will keep factors of $T_0$ explicit to facilitate keeping track of units.
The Kasner solutions (\ref{kasner}) can be analytically continued to static, anisotropic vacuum domain walls
by setting $T=ir$, $x^1=it$, and  $T_0=ir_0$. This yields the family of metrics
\begin{equation}\label{levic}
ds^2 =   dr^2- (r/r_0)^{2p_1}dt^2 + \sum_{i=2}^{3}(r/r_0)^{2p_i}(dx^i)^2.
\end{equation}
that is locally equivalent to the static, cylindrically symmetric vacuum solutions found by Levi-Civita \cite{levi-civita}, more
commonly written as
\begin{equation}\label{cylinder}
ds^2 = -(R/R_0)^{4\sigma}dt^2 +(R/R_0)^{4\sigma(2\sigma-1)}(dR^2+dz^2)+(R/R_0)^{-4\sigma}R^2d\varphi^2
\end{equation}
where $\varphi$ is an azimuthal angular coordinate with range $0\le\varphi <2\pi$.  The Levi-Civita parameter $\sigma$ takes values in the range $-\infty\le\sigma\le\infty$.  The metrics (\ref{levic}) and (\ref{cylinder}) are related through the radial coordinate transformation  $r/r_0=(R/R_0)^\lambda$ with 
%$\lambda=4\sigma^2-2\sigma+1$ 
%
\begin{equation}\label{lambdaparam}
\lambda=4\sigma^2-2\sigma+1
\end{equation}
and $r_0=R_0/(\lambda-1)$, and additionally setting $x^1=z$ and $x_2=R_0\varphi$.  The exponents in (\ref{levic}) are then found to be given in terms of the parameter $\sigma$ by
\begin{equation}\label{sigmaparam}
p_1={2\sigma\over \lambda}, \qquad p_2={2\sigma(2\sigma-1)\over\lambda},\qquad p_3={1-2\sigma\over \lambda},
\end{equation}
which  satisfy the Kasner constraints (\ref{constraints}).    We can think of the metric (\ref{cylinder}) as the Levi-Civita `fluxtube' and the metric (\ref{levic}) as the Levi-Civita `domain wall'.

The relations (\ref{sigmaparam}) give one parameterization of the solutions to the Kasner constraints.  
An alternative parameterization is given by an angle $\psi$ on 
the `Kasner circle' of radius ${2\over 3}$ centered at the point $({1\over 3},0)$ shown in Figure (\ref{kasner-circle}),  with the exponents given by
\begin{equation}\label{circleparam}
p_1={1\over 3}[1+ 2\cos(\psi+{4\pi\over 3})],\quad
p_2={1\over 3}[1+ 2\cos(\psi+{2\pi\over 3})],\quad
p_3={1\over 3}[1+ 2\cos(\psi)]
\end{equation}
Taking $\psi\rightarrow\psi+2\pi/3$ permutes the Kasner exponents and yields
physically equivalent spacetimes.   The Kasner angle $\psi$ and Levi-Civita parameter $\sigma$  are related according to
\begin{equation}
\cos\psi = {1-2\sigma-2\sigma^2\over 1-2\sigma+4\sigma^2}
\end{equation}

It will be useful to look at a few special cases of Kasner spacetimes.
The solutions with $\psi=0,2\pi/3,4\pi/3$ are Minkowski spacetime with Milne coordinates.   For example, taking $\psi=0$ gives
\begin{equation}\label{milne}
ds^2=-dT^2 +(dx^1)^2+(dx^2)^2+T^{2}(dx^3)^2
\end{equation}
while the other two choices for $\psi$ permute the $T^2$ factor to the other spatial coordinates.
The three possiblities correspond to $\sigma=0,{1\over 2},\pm\infty$ respectively.
We can also see from Figure (\ref{kasner-circle}) that with the exception of these flat solutions precisely one of the Kasner exponents is always negative.

Plane symmetric Kasner solutions, such that two exponents are equal, will also be important for us.   Up to permutation of the exponents there are two such solutions.  With $p_1=p_2$ these are $\psi=0,\pi$, or correspondingly $\sigma=0,1$.   As already noted, the $\psi=0$ solution is Minkowski spacetime.  The $\psi=\pi$ solution
\begin{equation}\label{nonflat}
ds^2=-dT^2 + T^{4/3}[(dx^1)^2+(dx^2)^2]+T^{-2/ 3}(dx^3)^2
\end{equation}
is a non-flat plane symmetric cosmology.    Equivalent solutions with different permutations of the exponents occur for $\psi=\pm\pi/3$.
Depending on precisely which two exponents are equal, the plane symmetric Kasner solutions  analytically continue to Levi-Civita spacetimes that have either planar rotational symmetry or a boost symmetry.  We will be primarily interested in this latter case, which corresponds to
the Levi-Civita fluxtube with $\psi=\pi$ ($\sigma=1$),  is given by
\begin{equation}\label{boost-sym}
ds^2 = (R/R_0)^4[-dt^2 +dR^2+dz^2]+(R/R_0)^{-4}R^2d\varphi^2
\end{equation}
and has a boost symmetry in the $z$-direction, but is non-flat.

\section{Interpolating via Melvin}\label{interpolating}

It is well known that  BPS black hole solutions such as the extreme Reisner-Nordstrom black hole interpolate between different fully supersymmetric vacuum states at infinity and the near horizon regime.  Melvin spacetimes, both the fluxtube (or domain wall) and cosmology, interpolate between solutions of the vacuum Einstein equations in a somewhat similar manner.  For the fluxtube (or domain wall)  this happens between the small and large radius limits, while for the cosmology it occurs between the limits of early and late times.

\begin{figure}\begin{center}
\includegraphics[width=0.7\textwidth]{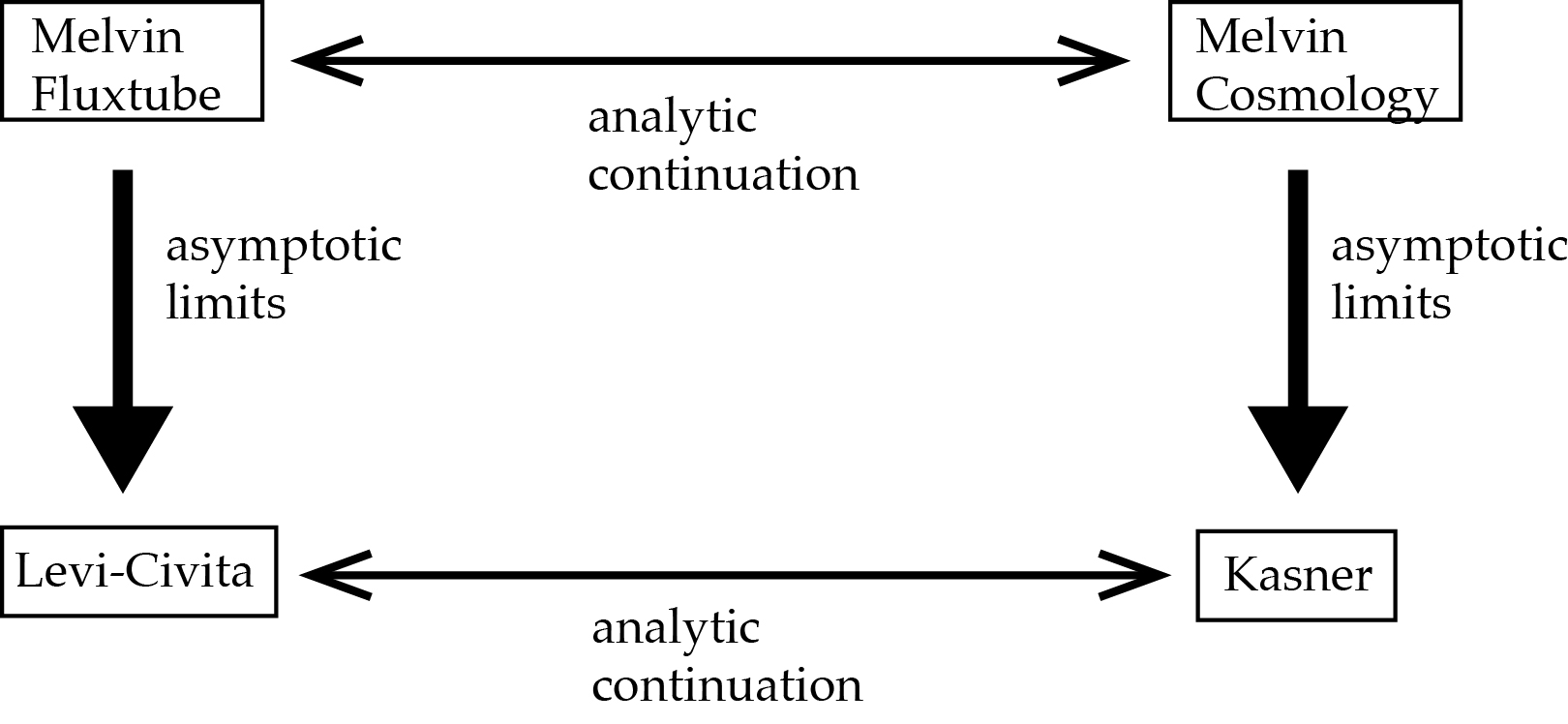}
\end{center}
\caption{{\small\sl A schematic display of  interpolating properties of Melvin fluxtubes and cosmologies.  Melvin cosmologies interpolated between different vacuum, anisotropic Kasner cosmologies in their early and late time limits, while Melvin fluxtubes interpolate at small and large radius between the corresponding vacuum, Levi-Civita solutions.}} 
\label{interpolatingfigure}
\end{figure}

Let us first consider the small radius limit, such that $Br\ll 1$,  of the Melvin solution in its fluxtube form (\ref{melvin}).  In this limit the function  $f\simeq 1$ and the metric is given approximately by the Minkowski metric in cylindrical spatial coordinates
\begin{equation}
ds^2\simeq dr^2-dt^2+dy^2+r^2 d\varphi^2,
\end{equation}
which we will choose to think of as the boost-symmetric Levi-Civita metric (\ref{cylinder})  with exponents $p_0=p_2=0$ and $p_3=1$, or equivalently 
parameter  $\sigma=0$.  Now consider the opposite, large radius limit of the Melvin fluxtube (\ref{melvin}) such that $Br\gg 1$.  We then have  $f\simeq B^2r^2/4$ and the metric  approaches
\begin{equation}
ds^2\simeq {B^4r^4\over 16}(dr^2-dt^2+dy^2) +{16\over B^4 r^2} d\varphi^2
\end{equation}
Comparing to equation  (\ref{boost-sym}), we see this is the Levi-Civita fluxtube with $\sigma=1$, which besides flat spacetime is the only other non-flat boost-symmetric Levi-Civita solution.  The Melvin fluxtube can then be thought of as interpolating between the two vacuum boost symmetric Levi-Civita spacetimes, with $\sigma=0$ and $\sigma=1$ respectively, in its small and large radius limits.  The same analysis holds for the domain wall forms of the Melvin and Levi-Civita solutions.

The analysis for the Melvin cosmology (\ref{cosmomelvin}) is essentially the same.  In this case, if we look in the early time limit, such that $ET\ll 1$, then the spacetime approaches Minkowski spacetime in Milne coordinates (\ref{milne}).  In the late time limit, with $ET\gg 1$, a transformation of the time coordinate shows that it approaches the non-flat, plane symmetric Kasner metric (\ref{nonflat}).  The Melvin cosmology, in either of its either electric or magnetic forms, can then be thought of as interpolating between the two plane symmetric vacuum Kasner spacetimes, with  $\sigma=0$ and $\sigma=1$ respectively, in the limits at early and late times.
The limiting behaviors for the Melvin fluxtube and cosmology are schematically illustrated in Figure (\ref{interpolatingfigure}).

These results raise further questions.  Is there a generalization of the Melvin cosmology (\ref{cosmomelvin}) that interpolates between an arbitrary Kasner spacetime  and some other Kasner spacetime in the early and late time limits?  If so, what is the map between the vacuum solutions at early and late times?  Correspondingly,  is there a generalization of the Melvin fluxtube (\ref{melvin}) that interpolates between an arbitrary Levi-Civita spacetime near the axis and a different one at infinity?  We will construct such generalized Melvin spacetimes in the next section.

\section{Generalized Melvin}\label{generating}

\begin{figure}
\begin{center}
\includegraphics[width=0.7\textwidth]{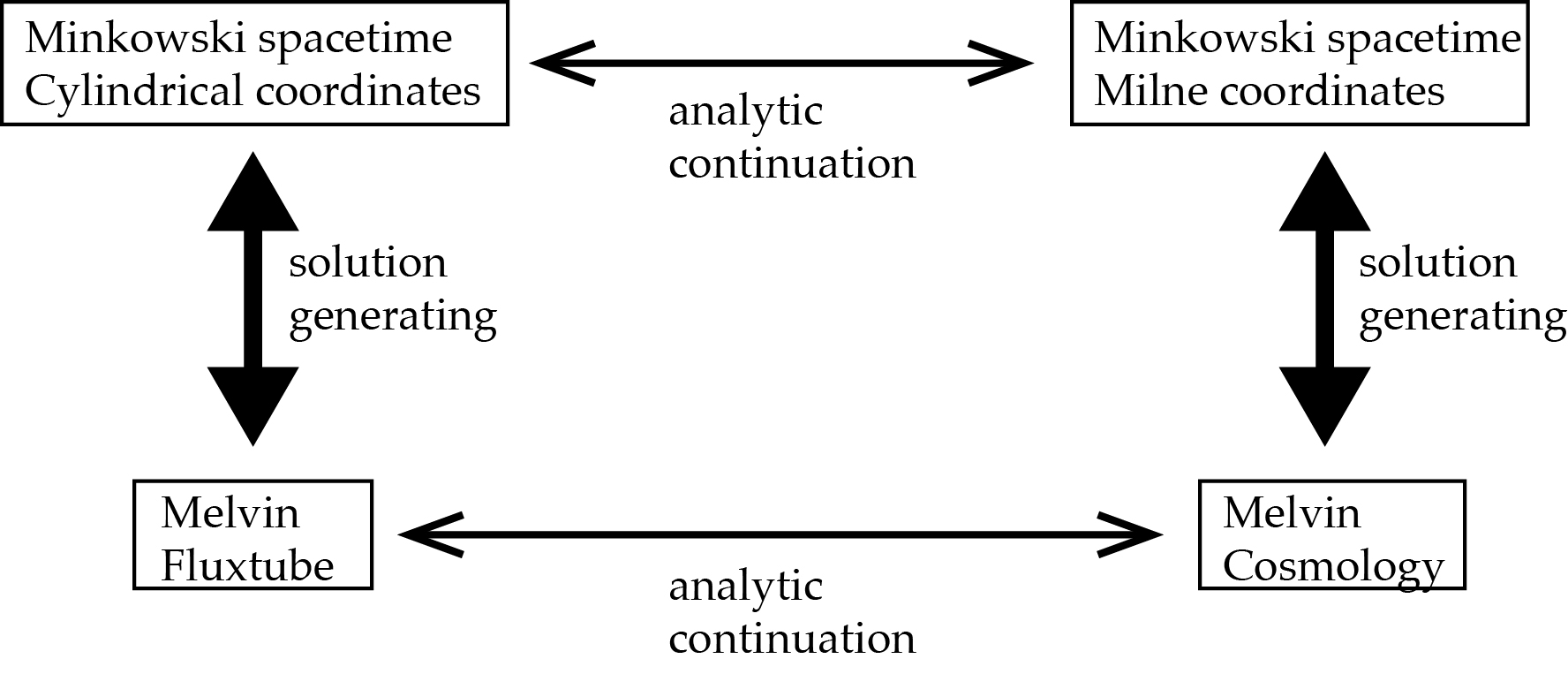}
\end{center}
\caption{{\small\sl A schematic display of  the relations via analytic continuation and solution generating between the Melvin fluxtube and cosmology, and Minkowski spacetime in different coordinate systems}} 
\label{generatingfigure}
\end{figure}

In order to construct appropriately generalized Melvin spacetimes, we recall how the Melvin fluxtube may be generated starting from Minkowki spacetime  \cite{harrison,Ernst3}.  Assume that a solution to the Einstein-Maxwell equations has a Killing vector $\partial/\partial\chi$ and denote the spacetime coordinates by $(y^i,\chi)$ with $i=1,2,3$, so that the spacetime metric and gauge potential are specified in terms of  $(g_{ij},g_{i\chi},g_{\chi\chi},A_i,A_\chi)$.   Further assume for simplicity that $g_{i\chi}=0$ and also that the gauge field of the seed solutions vanishes, so that $A_i=A_\chi=0$.   A new solution to the Einstein-Maxwell equations  is then given by
\begin{align}\label{transformation}
g_{ij}^\prime &= f^2 g_{ij},\qquad g_{\chi\chi}^\prime=f^{-2}g_{\chi\chi},\qquad A^\prime_\chi = {2\over Bf}\\
f  &= 1+{B^2\over 4} g_{\chi\chi}\nonumber
\end{align}
where $B$ is a free parameter and $g^\prime_{i\chi}=A^\prime_i=0$ \cite{harrison,Ernst3}.  Applying this transformation to Minkowski spacetime  in cylindrical coordinates, with $\chi$ taken to be the azimuthal coordinate,  yields the Melvin fluxtube (\ref{melvin}).  One similarly finds that applying it to Minkowski spacetime in Milne coordinates (\ref{milne}), with $\chi$ identified with the coordinate $x^3$,  yields the Melvin cosmology (\ref{cosmomelvin}).  This situation is illustrated in Figure (\ref{generatingfigure}).

Recall from the previous section that the Melvin fluxtube interpolates between flat spacetime at small radius and the non-flat, boost symmetric Levi-Civita spacetime (\ref{boost-sym}) at large radius.  We want to find a generalization of Melvin that interpolates between an arbitrary Levi-Civita fluxtube near the axis and another one at large radius.  This can be obtained by acting with the transformation (\ref{transformation}) on the general Levi-Civita fluxtube (\ref{cylinder}).  In this way one obtains the solution to the Einstein-Maxwell equations
\begin{align}\label{genflux}
ds^{\prime 2} &= f^2\left [ -(R/R_0)^{4\sigma}dt^2 +(R/R_0)^{4\sigma(2\sigma-1)}(dR^2+dz^2)\right] +f^{-2}(R/R_0)^{-4\sigma}R^2d\varphi^2  \\
A^\prime_\phi &= {2\over Bf},\qquad  f= 1+{B^2 R_0^2\over 4}(R/R_0)^{2-4\sigma}\nonumber
\end{align}
For $\sigma=0$ this reduces to the original Melvin fluxtube (\ref{melvin}), while for general values of  $\sigma$ it is a generalization of the Melvin fluxtube\footnote{See \cite{Kirezli:2012vw} and references therein for an account of the earlier history of these solutions.}.  These are not regular at $R=0$, but in the cosmological setting this singular behavior becomes the big bang.

We can carry out a similar construction starting from the general Kasner cosmologies (\ref{kasner}) to obtain generalized Melvin cosmologies.   Taking $\chi$-direction  in this case to be $x^3$-direction  and using $E$ for the parameter, we obtain the solution to the Einstein-Maxwell equations
\begin{align}\label{gencosmo}
ds^{\prime 2} &= f^2\left [ - dT^2+(T/ T_0)^{2p_1}(dx^1)^2+(T/ T_0)^{2p_2}(dx^2)^2\right ] +f^{-2}(T/ T_0)^{2p_3}(dx^3)^2\\   \nonumber
A_3^\prime &= {2\over Ef},\qquad f=1+{E^2\over 4}(T/ T_0)^{2p_3}
\end{align}
Starting from a general Kasner metric has given a generalized anisotropic Melvin anisotropic cosmology with an electric field along the $z$-direction.  Electromagnetic duality can by used to convert the electric field to a magnetic field.  In Appendix (\ref{rosensolns}) we show that the magnetic versions of these reproduce the Rosen solutions \cite{rosen1,rosen2}\footnote{A closely related spacetime also appeared recently as a fake-supersymmetric solution of Einstein gravity coupled to a gauge field with a wrong sign kinetic term \cite{Sabra:2015vca}.}.  Setting $p_3=1$ and $p_1=p_2=0$, corresponding to $\sigma=0$, this is the original Melvin cosmology (\ref{cosmomelvin}).   

\section{Interpolating map}\label{geninterpolating}

In this section we examine the limiting behaviors of the generalized Melvin fluxtubes and cosmologies found above, showing 
that they interpolate respectively between different vacuum Levi-Civita and Kasner limits.
This is most straightforward to see for the generalized Melvin fluxtubes (\ref{genflux}).  First assume that the Levi-Civita parameter is in the range $\sigma<1/2$. At sufficiently small radius we then have $f\simeq 1$ and the metric approaches the original Levi-Civita spacetime used as a seed for the solution generating map (\ref{transformation}).  At large radius, however, the second term in $f$ dominates and the metric, after linear rescaling of the coordinates,  has the form
\begin{equation}
ds^{\prime 2}\simeq - (R/R_0)^{4-4\sigma}dt^2 +(R/R_0)^{8\sigma^2-12\sigma+4}(dR^2+dz^2) +(R/R_0)^{-4+4\sigma}R^2d\phi^2
\end{equation}
We recognize this as the Levi-Civita fluxtube metric (\ref{cylinder}) with parameter $1-\sigma$.  
With $\sigma>1/2$, the two regions are reversed.  We find the original seed Levi-Civita spacetime parameterized by $\sigma$ at large radius and the new one parameterized by $1-\sigma$ at small radius.  In either case, the resulting interpolating map between Levi-Civita solutions in the small and large radius is given by $\sigma\rightarrow 1-\sigma$, generalizing the result for the original Melvin fluxtube (\ref{melvin}) which interpolates between  Levi-Civita solutions with $\sigma=0$ and $\sigma =1$.

The situation for the generalized Melvin cosmologies (\ref{gencosmo}) is analogous.  In this case we work in terms of the Kasner exponents.
Start by assuming that the exponent $p_3>0$  in (\ref{gencosmo}), so that the the function $f\simeq 1$ at early times and the spacetime approaches the seed Kasner metric with parameters $(p_1,p_2,p_3)$ in this limit.  At late times, however, the second term in $f$ is dominant and the metric approaches
\begin{equation}
ds^{\prime 2}\simeq {B^4\over 16} (T/T_0)^{4p_3}[- dT^2 +(T/T_0)^{2p_1}(dx^1)^2
+ (T/T_0)^{2p_2}(dx^2)^2] + {16\over B^4}(T/T_0)^{-2p_3}(dx^3)^2
\end{equation}
This can be put into Kasner form via the coordinate transformation $\tilde T/ \tilde T_0=  (T/T_0)^{2p_3+1}$ with $\tilde T_0=B^2 T_0/4(2p_3+1)$ and one finds that the new Kasner exponents are given by
\begin{equation}\label{newexponents}
\tilde p_1 = {p_1 +2p_3\over 2p_3+1},\qquad \tilde p_2 = {p_2 +2p_3\over 2p_3+1},\qquad \tilde p_3 = -{p_3\over 2p_3+1}
\end{equation}
which also follow from equations (\ref{lambdaparam}) and (\ref{sigmaparam}) by replacing $\sigma$ with $\tilde\sigma=1-\sigma$.  Hence the Melvin cosmologies, and as we will see also the scalar-Melvin cosmologies studied below, have behavior that is reminiscent of the BKL approach to a spacelike singularity, in which the universe bounces between different approximate Kasner regimes.

\section{Adding a massless scalar}\label{scalarsection}

Generalized Melvin fluxtubes and cosmologies with similar interpolating properties can also be constructed with an additional massless scalar field.
We will also include an exponential coupling of the scalar to the electromagnetic field and consider
the $4D$ dilaton gravity theory given by
\begin{equation}\label{action}
S=\int d^4x\sqrt{-g}\left(R-2(\nabla\phi)^2-e^{-2a\phi}F^2 \right),
\end{equation}
For dilaton coupling $a=0$ this reduces to Einstein-Maxwell theory minimally coupled to a massless scalar field, while for $a\neq 0$ the electromagnetic field acts as a source for the scalar field.  
In addition to the $a=0$ case, we will be particularly interested in $a=\sqrt{3}$ which arises from the Kaluza-Klein reduction of $5D$ Einstein gravity.
The theory with $a=1$ arises in string theory, while  dilaton coupling $a=1/\sqrt{3}$ comes from the dimensional reduction of $5D$ Einstein-Maxwell theory.  
One starts with the Melvin fluxtube solutions to dilaton gravity \cite{Gibbons:1987ps} given by
\begin{eqnarray}\label{scalarmelvin}
ds^2 &=& f^{{2\over 1+a^2}}(dr^2 -dt^2 +dy^2)+ f^{-{2\over 1+a^2}}r^2d\varphi^2,\qquad A_\varphi = {Br^2\over 2f}\\
e^{-2a\phi}&=&f^{{2a^2\over 1+a^2}},\qquad f=1+{(1+a^2)B^2r^2\over 4}\nonumber
\end{eqnarray}
For $a=0$, the scalar field vanishes and  this reduces to the Melvin fluxtube of Einstein-Maxwell theory (\ref{melvin}).  A notable feature of the dilaton Melvin fluxtubes is that for dilaton coupling $a<\sqrt{3}$ the size of constant $r$ circles contracts to zero for large $r$, giving it the `teardrop' shape of the original Melvin fluxtube (\ref{melvin}).  For $a>\sqrt{3}$, on the other hand, the size of the circles continues to increase at large radius, although less quickly than in flat spacetime.  For $a=\sqrt{3}$ the circles have asymptotically constant radius.

The dilaton Melvin fluxtube can be converted into a domain wall solution by setting $\varphi = Bz$ and taking $z$ to run from $-\infty$ to $\infty$.  The dilaton Melvin domain wall can then be analytically continued, precisely as in Section (\ref{melvinsection}), to yield anisotropic dilaton Melvin cosmologies, given by
\begin{eqnarray}\label{scalarcosmo}
ds^2 &=& f^{{2\over 1+a^2}}(-dT^2 +dy^2 +dy^2)+ f^{-{2\over 1+a^2}}E^2T^2dz^2,\qquad A_z= {E^2r^2\over 2f}\\
e^{-2a\phi}&=&f^{{2a^2\over 1+a^2}},\qquad f=1+{(1+a^2)E^2T^2\over 4}\nonumber
\end{eqnarray}
Whether the dilaton coupling $a$ is greater than, less than, or equal to the critical value $\sqrt{3}$ determines whether the scale factor in the $z$ direction is increasing, decreasing, or asymptotically constant in the late time limit.

We now  proceed in parallel with the previous sections by introducing  dilaton versions of Levi-Civita and Kasner spacetimes.  These will in turn be used to build generalizations of the dilaton Melvin fluxtube (\ref{scalarmelvin}) and cosmology (\ref{scalarcosmo}) that have interpolating properties analogous to those found above without the scalar field.
The dilaton Levi-Civita and Kasner spacetimes, like their non-dilatonic counterparts,  have vanishing electromagnetic field strength and hence are independent of the dilaton coupling.  In particular, the original Levi-Civita and Kasner spacetimes with a constant value for the scalar field are solutions to dilaton gravity (\ref{action}) for any value of the dilaton coupling.  However, there are also solutions with a nonconstant scalar field.  These can be obtained by focusing on the theory with $a=\sqrt{3}$ and making use of its equivalence to the dimensional reduction $5D$ Einstein gravity.

A solution to $a=\sqrt{3}$ dilaton gravity is obtained from a solution to $5D$ Einstein gravity, having translational invariance in the extra dimension, by identifying the $4D$ fields $(g_{ab}, A_a, \phi)$  in terms of the 
components of the $5D$ metric according to
\begin{equation}\label{reduce}
ds_5^2 = e^{-4\phi/\sqrt{3}}(dw +2A_a dx^a)^2 +e^{+2\phi/\sqrt{3}}g_{ab}dx^adx^b
\end{equation}
where $w$ is the additional spatial coordinate.  If we consider only solutions with  $A_a=0$, then the resulting metric and scalar field will be solutions to dilaton gravity for any value of $a$.
Let us start with the $5D$ Kasner spacetimes
\begin{equation}
ds_5^2 = -d\tildet^2 + \sum_{k=1}^3(\tildet/\tildet_0)^{2p_k}(dx^k)^2 + (\tildet/\tildet_0)^{2s}dw^2
\end{equation}
with exponents satisfying the $5D$ Kasner constraints
\begin{equation}\label{scalarconstraints}
\sum_{k=1}^3 p_k +s = \sum_{k=1}^3 p_k^2 +s^2=1
\end{equation}
Reducing to $4D$ using the decomposition (\ref{reduce}) of the $5D$ metric and then transforming to a new time coordinate that restores the Kasner form of the metric, one obtains the $4D$ scalar-Kasner solutions \cite{Belinski:1973zz}
\begin{equation}\label{scalarkasner}
ds^2 = - dT^2 +\sum_{i=1}^3 (T/ T_0)^{2\tilde p_i}(dx^i)^2,\qquad e^{{-4\phi\over\sqrt{3}}} =(T/T_0)^{2\tilde s}
\end{equation}
The new $4D$ scalar-Kasner exponents $(\tilde p_k,\tilde s)$ are  related to the original $5D$ Kasner exponents by
\begin{equation}\label{4dexponents}
\tilde p_k={2p_k+s\over 2+s},\qquad \tilde s = {2s\over 2+s}.
\end{equation}
It follows from the $5D$ Kasner constraints (\ref{scalarconstraints}) that these  exponents satisfy the new scalar-Kasner constraints
\begin{equation}\label{newconstraints}
\sum_{k=1}^3 \tilde p_k =1,\qquad 
\sum_{k=1}^3\tilde p_k^2  = 1 - {3\over 2} \tilde s^2
\end{equation}

Solutions to these constraints can be parameterized by the scalar exponent $\tilde s$ together with an angle $\psi$ on a `scalar-Kasner circle' whose radius 
depends on $\tilde s$.
The spatial exponents are given by
\begin{equation}\label{scalarcircleparam}
\tilde p_1={1\over 3}+ R\cos(\psi+{4\pi\over 3}),\quad
\tilde p_2={1\over 3}+ R\cos(\psi+{2\pi\over 3}),\quad
\tilde p_3={1\over 3}+ R\cos(\psi)
\end{equation}
and the radius by $R^2={4\over 9}-\tilde s^2$.
%%
%\begin{equation}
%R^2={4\over 9}-\tilde s^2.
%\end{equation}
%%
The scalar exponent must lie\footnote{For the $D=N+1$ dimensional Kasner constraints, one can show that all 
exponents lie in the range $-{N-2\over N}\le p\le 1$. For $D=5$ this implies that $-{1\over 2}\le s\le 1$, which using (\ref{4dexponents}) yields the range for $\tilde s$. } in the range $|\tilde s|\le 2/3$.  At the extremes of this range, with $\tilde s=\pm 2/3$,  the radius of the scalar-Kasner circle shrinks to zero.  The spatial Kasner exponents 
$\tilde p_k$ in this case are all equal to ${1\over 3}$, giving the well-known isotropic FRW solution for a massless scalar field
\begin{equation}\label{scalarFRW}
ds^2=-dT^2 +(T/T_0)^{2/3}[\sum_{k=1}^3 (dx^k)^2],\qquad e^{{-4\phi\over\sqrt{3}}} =(T/T_0)^{\pm{ 4\over 3}}.
\end{equation}
which corresponds to a perfect fluid with equation of state parameter $w=1$.  Note more generally from  (\ref{scalarcircleparam}), that in contrast to the vacuum Kasner solutions, if $\tilde s^2> {1\over 3}$, then all the the spatial Kasner exponents will be positive and the universe will be expanding in all directions.  Finally, the scalar-Kasner solutions may be analytically continued to obtain scalar-Levi-Civita spacetimes.  Letting $T=ir$, $x^1=it$ and $T_0=ir_0$ in (\ref{scalarkasner}) gives these in the static, anisotropic domain wall form
\begin{equation}
ds^2 = dr^2 -(r/r_0)^{2\tilde p_1} dt^2+\sum_{k=2}^3 (r/r_0)^{2\tilde p_k}(dx^k)^2,\qquad e^{{-4\phi\over\sqrt{3}}} =(r/r_0)^{2\tilde s}
\end{equation}

Electric and magnetic fields can now be added via solution generating to the scalar-Kasner and scalar-Levi-Civita solutions in order to obtain generalizations of the dilaton Melvin cosmologies and fluxtubes.  The solution generating technique used in Section (\ref{generating}) was generalized to dilaton gravity in \cite{Dowker:1993bt}. 
Assume that a solution to the equations of motion of dilaton gravity (\ref{action})  has a Killing vector $\partial/\partial\chi$ and again denote the spacetime coordinates by $(y^i,\chi)$ with $i=1,2,3$, so that now the spacetime  are specified in terms of  $(g_{ij},g_{i\chi},g_{\chi\chi},A_i,A_\chi,\phi)$.   Further assume for simplicity that $g_{i\chi}=0$ and also that the seed solution has $A_i=A_\chi=0$.   A new solution to dilaton gravity  is then given by
\begin{align}\label{dilatontransformation}
g_{ij}^\prime &= f^{2\over 1+a^2} g_{ij},\qquad g_{\chi\chi}^\prime=f^{-{2\over 1+a^2}}g_{\chi\chi},\qquad A^\prime_\chi = {2\over (1+a^2)Bf}\\
e^{-2a\phi^\prime} &= f^{2a^2\over 1+a^2}e^{-2a\phi},\qquad f  = 1+{(1+a^2)\over 4} B^2 g_{\chi\chi}e^{2a\phi}\nonumber
\end{align}
where $B$ is a free parameter and $g^\prime_{i\chi}=A^\prime_i=0$.  Applying this transformation to Minkowski spacetime  in cylindrical coordinates yields the dilaton Melvin fluxtube (\ref{scalarmelvin}) \cite{Dowker:1993bt,Dowker:1995gb}, while applying it to Minskoski spacetime in Milne coordinates yields the dilaton Melvin cosmology (\ref{scalarcosmo}).

Let us focus on dilaton cosmologies and begin by adding an electric field in the $x^3$-direction to the scalar-Kasner solutions (\ref{scalarkasner}).  For simplicity, let us also begin by working with the $a=0$ case, Einstein-Maxwell theory coupled to a massless scalar field.   Taking the limit $a\rightarrow 0$, the transformation (\ref{dilatontransformation}) reduces to the original transformation (\ref{transformation}) supplemented by the rule $\phi^\prime = \phi$ for the scalar field.
Taking $\chi$ in  (\ref{dilatontransformation})  to be the $x^3$-direction and denoting the parameter by $E$ we obtain the generalized scalar Melvin cosmologies
\begin{align}\label{genscalarcosmo}
ds^{\prime 2} &= f^2\left [ - dT^2+(T/ T_0)^{2\tilde p_1}(dx^1)^2+(T/ T_0)^{2\tilde p_2}(dx^2)^2\right ] +f^{-2}(T/ T_0)^{2\tilde p_3}(dx^3)^2\\   \nonumber
A_3^\prime &= {2\over Ef},\qquad e^{{-4\phi^\prime\over\sqrt{3}}} =(T/T_0)^{2\tilde s}, \qquad f=1+{E^2\over 4}(T/ T_0)^{2\tilde p_3}
\end{align}
If the scalar exponent $\tilde s =0$, these reduce to the original generalized Melvin cosmologies given in (\ref{gencosmo}).  For $\tilde s\neq 0$ we have the shifted exponents found in the scalar-Kasner solutions (\ref{scalarkasner}).

We can check that the generalized scalar Melvin cosmologies (\ref{genscalarcosmo}) interpolate between different scalar-Kasner limits.  Assume that $\tilde p_3> 0$, so that at early times $f\simeq 1$ and the metric approaches the original seed scalar-Kasner metric.  At late times, the function $f\simeq E^2(T/T_0)^{2\tilde p_3}/4$.  At first sight, the metric (\ref{genscalarcosmo}) in this limit does not have the scalar-Kasner form.  However, via a change in the time coordinate such that $T/T_0=(T^\prime/T_0^\prime)^{1\over 2\tilde p_3+1}$ the spacetime fields in the late time limit become
\begin{equation}
ds^{\prime 2}\simeq -dT^{\prime 2} +\sum_{k=1}^3 (T^\prime/T_0^\prime)^{2\tilde p_k^\prime}(dx^k)^2,\qquad
e^{{-4\phi^\prime\over\sqrt{3}}} =(T^\prime/T_0^\prime)^{2\tilde s^\prime}
\end{equation}
where the new scalar-Kasner exponents are given by
\begin{equation}\label{scalarinterp}
\tilde p^\prime_1 = {\tilde p_1 +2\tilde p_3\over 2\tilde p_3+1},\qquad \tilde p^\prime_2 = {\tilde p_2 +2\tilde p_3\over 2\tilde p_3+1},\qquad \tilde p^\prime_3 = -{\tilde p_3\over 2\tilde p_3+1},\qquad \tilde s^\prime ={\tilde  s\over 2\tilde p_3+1}
\end{equation}
It is easily checked that these satisfy the scalar-Kasner constraints (\ref{newconstraints}).
We see that the interpolating effect of the electric field on the spacetime exponents in the presence of the dilaton matches the effect on the spatial exponents when connecting two vacuum Kasner spacetimes in (\ref{newexponents}).  Note that the scalar exponent is modified between early and late times, even though the scalar field is unaffected by the introduction of the electric field, due to the transformation of the time coordinate needed to put the late time metric into the Kasner form\footnote{It would be nice to find a parameterization of solutions to the scalar-Kasner constraints such that  the interpolating map (\ref{scalarinterp}) has a simple form, similar to the map $\sigma\rightarrow 1-\sigma$ that holds in the non-scalar case.}.  As an example, consider starting with the FRW case (\ref{scalarFRW})  which has $\tilde s= {2\over 3}$ and $\tilde p_k ={1\over 3}$.  Adding an electric field in the $x^3$-direction breaks the isotropy and leads to a late time scalar-Kasner phase with exponents $\tilde s^\prime = {2\over 5}$, $\tilde p_1^\prime=\tilde p_2^\prime ={3\over 5}$ and $\tilde p_3^\prime = -{1\over 5}$.  Although all the spatial directions start out expanding at early times, we see that the $x^3$-direction is contracting in the late time scalar-Kasner phase.

Finally, all this can also be done including a nonzero dilaton coupling.  The generalized dilaton Melvin cosmology is then found to be
\begin{align}\label{gendilatoncosmo}
ds^{\prime 2} &= f^{2\over 1+a^2}\left [ - dT^2+(T/ T_0)^{2\tilde p_1}(dx^1)^2+(T/ T_0)^{2\tilde p_2}(dx^2)^2\right ] +f^{-{2\over 1+a^2}}(T/ T_0)^{2\tilde p_3}(dx^3)^2\\   \nonumber
A_3^\prime &= {2\over Ef},\qquad e^{-2a\phi^\prime} =f^{2a^2\over 1+a^2}(T/T_0)^{\sqrt{3} a\tilde s}, 
\qquad f=1+{1\over 1+a^2}E^2(T/ T_0)^{2\tilde p_3-\sqrt{3}a\tilde s}
\end{align}
If we assume that $\lambda\equiv {2\tilde p_3-\sqrt{3}a\tilde s\over 1+a^2}>0$, then at early times the spacetime is asymptotic to the seed solution with exponents $(\tilde p_k,\tilde s)$ while at late times it approaches the scalar-Kasner spacetime with exponents
\begin{equation}\label{dilatoninterp}
\tilde p^\prime_1 = {\tilde p_1 +\lambda\over \lambda +1},\qquad \tilde p^\prime_2 = {\tilde p_2 +\lambda\over \lambda+1},
\qquad \tilde p^\prime_3 = {\tilde p_3-\lambda\over \lambda +1},\qquad \tilde s^\prime ={\tilde  s+2a\lambda/\sqrt{3}\over \lambda +1}\, .
\end{equation}
These solutions may also be analytically continued to obtain generalized dilaton Melvin domain-wall/fluxtube solutions.

\section{Conclusions}\label{discussionsection}

In this paper we have explored different aspects of the duality, via analytic continuation, between Melvin fluxtubes and anisotropic electromagnetic cosmologies.  We have seen how solution generating techniques, used to construct fluxtube solutions, can also be used to construct the corresponding cosmologies, which have more commonly been found via brute force methods.  We have also seen how the interpolating feature of the cosmologies, between vacuum Kasner limits, has a counterpart in the radial evolution of Melvin fluxtubes between different vacuum Levi-Civita limits.  These observations strengthen our understanding of these classes of spacetimes by exposing an underlying common structure.  Finally, we have seen that a similar structure exists when a massless dilaton field is included with exponential coupling to the gauge field strength.

A number of interesting directions exist for further investigation.  One possibility is to look at the cosmological duals and interpolating properties of more general fluxbrane solutions \cite{Gutperle:2001mb}, including supersymmetric examples.  A second possibility is to add in a potential for the scalar field to understand the structure of  anisotropic inflationary models (see {\it e.g.} the review \cite{Maleknejad:2012fw}) and their fluxbrane counterparts.    This would hopefully yield a more systematic understanding of violations of the cosmic no-hair property \cite{Wald:1983ky} in these models.   A third direction is adding a cosmological constant, with potential (A)dS/CFT applications.

\appendix

\subsection*{Acknowledgements}
The authors thank Brian Harvie and Jason Stevens for helpful conversations.

\section*{Appendix}

\section{Rosen magnetic cosmologies}\label{rosensolns}

In this appendix we show that the generalized Melvin cosmologies (\ref{gencosmo}) we have constructed are equivalent to anisotropic cosmological solutions to the Einstein-Maxwell equations found many years ago by Rosen \cite{rosen1,rosen2}.  Electric versions of the Rosen solutions are given by
\begin{align}\label{rosensolutions}
ds^2 &= -{b^2 (\tan {t\over 2})^{2(c_1+c_2)}\over \sin^4 t}dt^2 
+{ (\tan{t\over 2})^{2c_1}\over\sin^2 t}(dy^1)^2 +{ (\tan {t\over 2})^{2c_2}\over\sin^2 t}(dy^2)^2 
+\sin^2 t (dy^3)^2\\
A & = {1\over \cos^2{t\over 2}} dy^3\nonumber
\end{align}
where the product $c_1c_2=1$.  The original magnetic versions can be recovered by electromagnetic duality.  The first step in demonstrating the equivalence of these with the generalized Melvin cosmologies is to set $\tau={t\over 2}$ and make use of trigonometric identities to obtain
\begin{align}\label{rosensolutions_v2}
ds^2 &= -{b^2 (\sin^2\tau)^{c_1+c_2-2}\over 4(\cos^2\tau)^{c_1+c_2+2}}    d\tau ^2 
+{(\sin^2\tau)^{c_1-1}\over 4(\cos^2\tau)^{c_1+1}} (dy^1)^2 +{(\sin^2\tau)^{c_2-1}\over 4(\cos^2\tau)^{c_2+1}} (dy^2)^2 
+4\sin^2\tau\cos^2\tau (dy^3)^2\\
A & = {1\over \cos^2\tau} dy^3\nonumber
\end{align}
Let us now make a further change in time coordinate such that 
\begin{equation}
\cos^2\tau = {1\over f},\qquad \sin^2\tau={1\over f}{E^2\over 4}(T/T_0)^{2\eta}
\end{equation}
with $f=1+{E^2\over 4}(T/T_0)^{2\eta}$ and  $\eta=(c_1+c_2-1)^{-1}$.  The metric and gauge field then have the form of the generalized Melvin cosmologies (\ref{gencosmo})  with
\begin{equation}
p_1=(c_1-1)\eta,\qquad p_2=(c_2-1)\eta,\qquad p_3=\eta
\end{equation}
if we take the dimensionful constant $b$ in (\ref{rosensolutions}) to be given according to
\begin{equation}
b^2 = {4\over \eta^2}\left ({E^2\over 4}\right )^{-{1\over \eta}} T_0^2
\end{equation}
%

%Rosen found planar symmetric Bianchi I solutions to the Einstein-Maxwell equations back in 1962 , with metrics given by
%%
%\begin{equation}
%ds^2=-{b^2 \over (1\pm\cos t)^4}dt^2 +{1\over (1\pm\cos t)^2}(du^2+dv^2)+\sin^2 tdw^2
%\end{equation}
%%
%
%We can show that these are equivalent to the simple magnetic cosmologies of section (\ref{simple}).


\begin{thebibliography}{99}


%\cite{Skenderis:2006jq}
\bibitem{Skenderis:2006jq} 
  K.~Skenderis and P.~K.~Townsend,
  ``Hidden supersymmetry of domain walls and cosmologies,''
  Phys.\ Rev.\ Lett.\  {\bf 96}, 191301 (2006)
  [hep-th/0602260].
  %%CITATION = HEP-TH/0602260;%%
  %87 citations counted in INSPIRE as of 16 Jan 2015
  
  %\cite{Skenderis:2006rr}
\bibitem{Skenderis:2006rr} 
  K.~Skenderis and P.~K.~Townsend,
  ``Hamilton-Jacobi method for curved domain walls and cosmologies,''
  Phys.\ Rev.\ D {\bf 74}, 125008 (2006)
  [hep-th/0609056].
  %%CITATION = HEP-TH/0609056;%%
  %37 citations counted in INSPIRE as of 16 juil. 2015
  
  %\cite{Skenderis:2006fb}
\bibitem{Skenderis:2006fb} 
  K.~Skenderis and P.~K.~Townsend,
  ``Pseudo-Supersymmetry and the Domain-Wall/Cosmology Correspondence,''
  J.\ Phys.\ A {\bf 40}, 6733 (2007)
  [hep-th/0610253].
  %%CITATION = HEP-TH/0610253;%%
  %53 citations counted in INSPIRE as of 16 juil. 2015
  
  %\cite{Melvin:1963qx}
\bibitem{Melvin:1963qx} 
  M.~A.~Melvin,
  ``Pure magnetic and electric geons,''
  Phys.\ Lett.\  {\bf 8}, 65 (1964).
  %%CITATION = PHLTA,8,65;%%
  %236 citations counted in INSPIRE as of 16 Jan 2015
  
  
    \bibitem{rosen1}
  G.~Rosen, ``Symmetries of the Einstein-Maxwell Equations," J. Math. Phys. {\bf 3}, 313 (1962).
  
  \bibitem{rosen2}
  G.~Rosen, ``Spatially Homogeneous Solutions to the Einstein-Maxwell Equations," Phys. Rev. {\bf 136}, B297 (1964).

  
   %\cite{harrison}
\bibitem{harrison}
B.K. Harrison, ``New solutions of the Einstein-Maxwell equations from old," J. Math. Phys. {\bf 9}, 1744 (1968)


 
 %\cite{Ernst3}
  \bibitem{Ernst3}
F. J. Ernst,
``Black holes in a magnetic universe,''
J. Math. Phys. {\bf 17}, 54 (1976)


  %\cite{Belinski:1973zz}
\bibitem{Belinski:1973zz} 
  V.~A.~Belinski and I.~M.~Khalatnikov,
  ``Effect of Scalar and Vector Fields on the Nature of the Cosmological Singularity,''
  Sov.\ Phys.\ JETP {\bf 36}, 591 (1973).
  %%CITATION = SPHJA,36,591;%%
  %135 citations counted in INSPIRE as of 29 Jun 2015
  
  %\cite{Belinsky:1970ew}
\bibitem{Belinsky:1970ew} 
  V.~A.~Belinsky, I.~M.~Khalatnikov and E.~M.~Lifshitz,
  ``Oscillatory approach to a singular point in the relativistic cosmology,''
  Adv.\ Phys.\  {\bf 19}, 525 (1970).
  %%CITATION = ADPHA,19,525;%%
  %629 citations counted in INSPIRE as of 16 juil. 2015
  
  %\cite{Gibbons:1986wg}
\bibitem{Gibbons:1986wg} 
  G.~W.~Gibbons and D.~L.~Wiltshire,
  ``Space-Time as a Membrane in Higher Dimensions,''
  Nucl.\ Phys.\ B {\bf 287}, 717 (1987)
  [hep-th/0109093].
  %%CITATION = HEP-TH/0109093;%%
  %285 citations counted in INSPIRE as of 27 Aug 2015
  
  %\cite{Wiltshire:1987ch}
\bibitem{Wiltshire:1987ch} 
  D.~L.~Wiltshire,
  ``Global Properties of {Kaluza-Klein} Cosmologies,''
  Phys.\ Rev.\ D {\bf 36}, 1634 (1987).
  %%CITATION = PHRVA,D36,1634;%%
  %36 citations counted in INSPIRE as of 27 Aug 2015

    
  %\cite{Miguelote:2000qi}
\bibitem{Miguelote:2000qi} 
  A.~Y.~Miguelote, M.~F.~A.~da Silva, A.~Wang and N.~O.~Santos,
  %``Levi-Civita solutions coupled with electromagnetic fields,''
  Class.\ Quant.\ Grav.\  {\bf 18}, 4569 (2001)
  [gr-qc/0104018].
  %%CITATION = GR-QC/0104018;%%
  %15 citations counted in INSPIRE as of 16 juil. 2015
  
  
   %\cite{Baykal:2005pv}
\bibitem{Baykal:2005pv} 
  A.~Baykal and O.~Delice,
  ``Cylindrically symmetric-static Brans-Dicke-Maxwell solutions,''
  gr-qc/0512143.
  %%CITATION = GR-QC/0512143;%%
  %1 citations counted in INSPIRE as of 16 Jul 2015

     
  %\cite{Kirezli:2012vw}
\bibitem{Kirezli:2012vw} 
  P.~Kirezli, D.~K.~'iftci and ….~Delice,
  ``Higher Dimensional Cylindrical or Kasner Type Electrovacuum Solutions,''
  Gen.\ Rel.\ Grav.\  {\bf 45}, 2251 (2013)
  [arXiv:1205.5336 [gr-qc]].
  %%CITATION = ARXIV:1205.5336;%%
  %1 citations counted in INSPIRE as of 16 juil. 2015
  
  %\cite{Kastor:2013nha}
\bibitem{Kastor:2013nha} 
  D.~Kastor and J.~Traschen,
  ``Magnetic Fields in an Expanding Universe,''
  arXiv:1312.4923 [hep-th].
  %%CITATION = ARXIV:1312.4923;%%

   \bibitem{levi-civita}
  T.~Levi-Civita, Rend. Acc. Lincei {\bf 26}, 307 (1917).

  
  %\cite{Kasner:1921zz}
\bibitem{Kasner:1921zz} 
  E.~Kasner,
  ``Geometrical theorems on Einstein's cosmological equations,''
  Am.\ J.\ Math.\  {\bf 43}, 217 (1921).
  %%CITATION = AJMAA,43,217;%%
  %153 citations counted in INSPIRE as of 19 Jan 2015
  
    %\cite{Sabra:2015vca}
\bibitem{Sabra:2015vca} 
  W.~A.~Sabra,
  ``Phantom Metrics With Killing Spinors,''
  arXiv:1507.04597 [hep-th].
  %%CITATION = ARXIV:1507.04597;%%



 %\cite{Gibbons:1987ps}
\bibitem{Gibbons:1987ps} 
  G.~W.~Gibbons and K.~-i.~Maeda,
  ``Black Holes and Membranes in Higher Dimensional Theories with Dilaton Fields,''
  Nucl.\ Phys.\ B {\bf 298}, 741 (1988).
  %%CITATION = NUPHA,B298,741;%%
  %868 citations counted in INSPIRE as of 15 Oct 2013


    
  %\cite{Dowker:1993bt}
\bibitem{Dowker:1993bt} 
  F.~Dowker, J.~P.~Gauntlett, D.~A.~Kastor and J.~H.~Traschen,
  ``Pair creation of dilaton black holes,''
  Phys.\ Rev.\ D {\bf 49}, 2909 (1994)
  [hep-th/9309075].
  %%CITATION = HEP-TH/9309075;%%
  %191 citations counted in INSPIRE as of 12 Feb 2015
  
   %\cite{Dowker:1995gb}
\bibitem{Dowker:1995gb} 
  F.~Dowker, J.~P.~Gauntlett, G.~W.~Gibbons and G.~T.~Horowitz,
  ``The Decay of magnetic fields in Kaluza-Klein theory,''
  Phys.\ Rev.\ D {\bf 52}, 6929 (1995)
  [hep-th/9507143].
  %%CITATION = HEP-TH/9507143;%%
  %131 citations counted in INSPIRE as of 27 gen 2015

  
  
  
  
 
  
    %\cite{Gutperle:2001mb}
\bibitem{Gutperle:2001mb} 
  M.~Gutperle and A.~Strominger,
  ``Fluxbranes in string theory,''
  JHEP {\bf 0106}, 035 (2001)
  [hep-th/0104136].
  %%CITATION = HEP-TH/0104136;%%
  %112 citations counted in INSPIRE as of 17 Oct 2013
  
%\cite{Maleknejad:2012fw}
\bibitem{Maleknejad:2012fw} 
  A.~Maleknejad, M.~M.~Sheikh-Jabbari and J.~Soda,
  ``Gauge Fields and Inflation,''
  Phys.\ Rept.\  {\bf 528}, 161 (2013)
  [arXiv:1212.2921 [hep-th]].
  %%CITATION = ARXIV:1212.2921;%%
  %71 citations counted in INSPIRE as of 16 juil. 2015
  
%\cite{Wald:1983ky}
\bibitem{Wald:1983ky} 
  R.~M.~Wald,
  ``Asymptotic behavior of homogeneous cosmological models in the presence of a positive cosmological constant,''
  Phys.\ Rev.\ D {\bf 28}, 2118 (1983).
  %%CITATION = PHRVA,D28,2118;%%
  %435 citations counted in INSPIRE as of 17 Jul 2015


   
\end{thebibliography}
\end{document}